\documentclass[aps,prl,reprint]{revtex4-1}

\usepackage{amsmath}
\usepackage{amssymb}
\usepackage{amsfonts}

\begin{document}

\title{Comment on ``The QCD axion beyond the classical level: A lattice study''}

\author{Ch.~Hoelbling}
\author{A.~Pasztor}
\affiliation{Department of Physics, University of Wuppertal, Gaussstr. 20, D-42119 Wuppertal, Germany}

\begin{abstract}
We rebut the claim by Nakamura and Schierholz \cite{Nakamura:2018pxj} that the mass of a potential axion needs
to be no less than $\sim 230 \mathrm{MeV}$ pointing out errors in both
their analytic argument and numerical simulations.
\end{abstract}

\maketitle

In a recent preprint \cite{Nakamura:2018pxj}, Nakamura and Schierholz claim that the standard relation for the axion mass $m_a$ being inversely proportional to the axion decay constant $f_a$ is wrong and that an axion mass of $m_a \lesssim 230 \mathrm{MeV}$ is ruled out based on lattice simulations of the Peccei-Quinn theory. In this short note, we point out a mistake in their analytic argument as well as crucial shortcomings of their lattice simulations that invalidates their claims.

Following the notation of \cite{Nakamura:2018pxj}, the axion action is
\begin{equation}
S=S_\mathrm{QCD}+\int d^4x \left(
\frac{1}{2} (\partial_\mu\phi_a(x))^2+i\frac{\phi_a(x)}{f_a}q(x)
\right)
\label{eq:action}
\end{equation}
where the topological charge density $q(x)$ is given by
\begin{equation}
q(x)=-\frac{1}{64\pi^2}\epsilon_{\alpha\beta\mu\nu}F_{\alpha\beta}^aF_{\mu\nu}^a
\end{equation}
The standard relation for the axion mass \cite{Bardeen:1986yb,Peccei:2006as} is
\begin{equation}
m_a^2=\frac{\partial^2}{\partial \bar{\phi}_a^2}\left. U_\mathrm{eff}(\bar{\phi}_a)\right|_{\bar{\phi}_a=0}=\frac{\chi_t}{f_a^2}
\end{equation}
where $U_\mathrm{eff}$ is the quantum effective potential depending on the mean field $\bar{\phi}_a=\int d^4x \phi_a(x)/V$ and $\chi_t$ denotes the topological susceptibility
\begin{equation}
\chi_t=\frac{\langle Q^2\rangle}{V}
\qquad
Q=\int d^4x q(x)
\end{equation}
The authors of \cite{Nakamura:2018pxj} note that by rescaling the axion field $\tilde\phi_a=\phi_a/f_a$ the second term of (\ref{eq:action}) becomes
\begin{equation}
\frac{1}{f_a^2}\int d^4x \left(
\frac{1}{2} (\partial_\mu\tilde\phi_a(x))^2+i\tilde\phi_a(x)q(x)
\right)
\end{equation}
and thus they claim that the axion mass does not depend on $f_a$ explicitly, but only through the topological charge density.

At this point it is important that the prefactor of the kinetic term is no more unity, which implies that the prefactor of the quadratic term of the effective potential in the rescaled field $\tilde\phi$ is not the pole mass but needs to be renormalized by exactly the factor $1/f_a$, invalidating their argument.

The argument in \cite{Nakamura:2018pxj} starts with the asymptotic behaviour of the correlation function of the original axion field  $\phi_a$ with a pseudoscalar source $\pi$ for large temporal separations
\begin{equation}
\int d^3x 
\langle
\phi_a(\vec{x},t)\pi(0)
\rangle
\simeq
Ae^{-m_at}
\end{equation}
and continues with the equation of motion
\begin{equation}
\frac{\partial^2}{\partial t^2}\int d^3x \langle
\phi_a(\vec{x},t)\pi(0)
\rangle
=
\frac{1}{f_a}
i\int d^3x
\langle
q(\vec{x},t)\pi(0)
\rangle
\qquad
t>0
\label{eq:prop}
\end{equation}
We note that in order for this equation to hold it is necessary that
\begin{equation}
\int d^3x 
\langle
 q(\vec{x},t)\pi(0)
\rangle
\simeq
Be^{-m_at}
\end{equation}
with $B=-i f_am_a^2A$. The argument in \cite{Nakamura:2018pxj} proceeds by rewriting (\ref{eq:prop}) in terms of the rescaled field $\tilde\phi_a$
\begin{equation}
\frac{\partial^2}{\partial t^2}\int d^3x \langle
\tilde\phi_a(\vec{x},t)\pi(0)
\rangle
=
i\int d^3x
\langle
q(\vec{x},t)\pi(0)
\rangle
\qquad
t>0
\label{eq:tildprop}
\end{equation}
which the authors claim implies that there is no explicit dependence of $m_a$ on $f_a$. With the ansatz
\begin{equation}
\int d^3x 
\langle
\tilde\phi_a(\vec{x},t)\pi(0)
\rangle
\simeq
\tilde Ae^{-m_at}
\end{equation}
we obtain $B=-i m_a^2\tilde A$. But since $\tilde A=f_a A$, (\ref{eq:prop}) and (\ref{eq:tildprop}) unsurprisingly have exactly the same content and it is not possible to deduce from them the $f_a$ dependence of the axion mass.

Ref. \cite{Nakamura:2018pxj} also uses lattice simulations to support
their claim of a light axion being impossible. We would also like to
point out a shortcoming of this study. 
   Ref. \cite{Nakamura:2018pxj} uses an expansion of
   the axion action in $\phi_a$, assuming a small value of the field $\phi_a$ at
   every lattice point. They claim this is a good approximation for
   large volumes. This is incorrect, only the average value $\bar
   \phi_a$ will be small, not the field value $\phi_a$ at every point. 
   This procedure results in an action with a scalar field of bare
   mass zero,
   with a simple Yukawa coupling to the quarks. This is a quite
   different theory then the Peccei-Quinn theory they claim
   to simulate. Most importantly, the lattice action used in
   \cite{Nakamura:2018pxj} has
   no symmetry protecting $m_a=0$. Consequently a counterterm of the
   form $m_a^2 \phi_a^2$ would need to be introduced into
   the lattice action and tuned in a standard fashion (like the
   hopping parameter has to be tuned for Wilson fermions because they
   break chiral symmetry). It is no
   surprise that without such a tuning a large mass is observed, since
    the soft breaking via
   $1/f_a$ is probably dwarfed by the explicit breaking in the action.
Therefore we believe these numerical studies are also inconclusive.

\end{document}